\documentclass[prb,twocolumn,showpacs,preprintnumbers,amsmath,amssymb,floatfix]{revtex4}

\usepackage[]{graphicx}         	
\usepackage{bm}             		
\usepackage{tabularx}

\begin{document}

%
%
\title{Temperature dependence of nonlinear auto-oscillator linewidths: Application to spin-torque nano-oscillators}

\author{V. S. Tiberkevich}
\email{tyberkev@oakland.edu}
\author{A. N. Slavin}
\email{slavin@oakland.edu}
\affiliation{Department of Physics, Oakland University, Rochester, MI 48309, USA}
\author{Joo-Von Kim}
\email{joo-von.kim@u-psud.fr}
\affiliation{Institut d'Electronique Fondamentale, UMR CNRS 8622, Universit{\'e} Paris-Sud, 91405 Orsay, France}

\date{\today}

%
%
\begin{abstract}

The temperature dependence of the generation linewidth for an auto-oscillator with a nonlinear frequency shift is calculated.  It is shown that the frequency nonlinearity creates a finite correlation time, $\tau= 1/\Gamma_{\rm p}$, for the phase fluctuations. In the low-temperature limit in which the spectral linewidth is smaller than $\Gamma_{\rm p}$, the line shape is approximately Lorentzian and the linewidth is linear in temperature. In the opposite high-temperature limit in which the linewidth is larger than $\Gamma_{\rm p}$, the nonlinearity leads to an apparent ``inhomogeneous broadening" of the line, which becomes Gaussian in shape and has a square-root dependence on temperature. The results are illustrated for the spin-torque nano-oscillator.

\end{abstract}


\maketitle

%
%

Spin-torque auto-oscillators (STOs), which are based on the spin-transfer effect in magnetic multilayered nanostructures,~\cite{Slonczewski:JMMM:1996,Berger:PRB:1996} possess a much stronger intrinsic frequency nonlinearity compared to conventional auto-oscillatory systems.~\cite{Lax:PR:1967,Risken:1989} For example, variations in the generation frequency can be two orders of magnitude greater than the generation linewidth in STOs, while this ratio is typically of the order of unity for conventional auto-oscillators. One important and interesting consequence of this nonlinearity is the strong coupling between the phase and amplitude fluctuations in the presence of thermal noise. While the amplitude fluctuations are subject only to a Gaussian white noise forcing, they act, through the nonlinear coupling, as a colored-noise source for the phase fluctuations.~\cite{Kim:PRL:2008,Kim:PRL:2008b} This coupling has been shown to cause substantial broadening of the generation linewidth above the generation threshold~\cite{Kim:PRL:2008} and to lead to asymmetric line shapes near threshold.~\cite{Kim:PRL:2008b}

In this article, we investigate the temperature dependence of the generation linewidth of a nonlinear oscillator using a universal model of an auto-oscillator with a strong nonlinear frequency shift.~\cite{Tiberkevich:APL:2007,Kim:PRL:2008,Kim:PRL:2008b} We demonstrate that the frequency nonlinearity creates a finite correlation time, $ \tau = 1/\Gamma_{\rm p}$, for the phase fluctuations caused by the thermal noise, which leads to a non-Lorentzian shape of the power spectrum in general. A Lorentzian line shape is recovered only in the limit of low temperatures (at which linewidth magnitudes $\Delta\omega$ are small compared to the inverse correlation time of the phase fluctuations $ \Delta\omega \ll \Gamma_{\rm p}$). The generation linewidth in this regime is linearly proportional to temperature. In the opposite limit of high temperatures (at which linewidth magnitudes are large compared to the inverse correlation time $\Delta\omega \gg \Gamma_{\rm p}$), the frequency nonlinearity leads to the apparent ``inhomogeneous broadening" of the linewidth in which fluctuations in the oscillation amplitude leads to variations of the generated frequency by virtue of the strong frequency nonlinearity. The line shape in this limit is Gaussian with a linewidth that is proportional to the square-root of temperature. The results are illustrated for a spin-torque auto-oscillator.

We study the stochastic dynamics within the framework of a universal nonlinear oscillator model with a stochastic force $f_n$ that represents the action of thermal fluctuations,~\cite{Kim:PRB:2006,Tiberkevich:APL:2007,Kim:PRL:2008,Kim:PRL:2008b,Slavin:ITM:2005}
\begin{equation}\label{model}
	\frac{dc}{dt} + i \omega(p)c + \Gamma_+(p)c - \Gamma_-(p)c = f_n(t),
\end{equation}
where $c = |c|\exp(i\phi)$ is a dimensionless complex variable that describes the excited mode amplitude $|c|$ and phase $\phi$, and $p=|c|^2$ is the dimensionless oscillation power. The equation of motion contains three terms which are functions of power: the nonlinear generation frequency $\omega(p)$, the positive nonlinear damping $\Gamma_+(p)$ determined by the natural dissipative processes in the system, and the negative nonlinear damping $\Gamma_-(p)$ determined by the active element that supplies external energy in the system. The stochastic term $f_n(t)$ describes the action of thermal fluctuations on the auto-oscillator and its statistical properties are determined from the thermodynamic properties of the oscillator at thermal equilibrium. This is achieved by taking $f_n(t)$ to be a white Gaussian noise with zero mean, $\langle f_n(t) \rangle = 0$, with a second-order correlation function, $ \langle f_n(t) f_n^*(t') \rangle = 2 D_n \delta(t-t')$, where the diffusion constant for a nonlinear oscillator is given by~\cite{Tiberkevich:APL:2007}
\begin{equation}
D_n(p) = \Gamma_+(p) \eta(p) = \Gamma_+(p) \frac{k_B T}{\lambda \omega(p)}.
\end{equation}
Here $\eta$ is the effective noise power in the nonlinear regime and $\lambda$ is a coefficient that relates the oscillator power $p$ to the oscillator energy $\mathcal{E}$.

In the following, we use the STO as an example of the stochastic nonlinear auto-oscillator described by the above model. At reasonably small generation powers, $p<1$, the coefficients of the power-dependent terms in Eq.~(\ref{model}) can be approximated with the following Taylor series expansions:~\cite{Kim:PRL:2008,Kim:PRL:2008b} $\omega(p) = \omega_0 + N p$,  $\Gamma_+(p) = \Gamma_0 (1 + Q p)$, and $\Gamma_-(p) = \sigma I (1-p)$, where $I$ is the bias current and the explicit definitions of the coefficients $N, Q,\sigma, \Gamma_0 $ are given in Ref.~\onlinecite{Kim:PRL:2008b}. The coefficient of proportionality between the energy and power of the STO is $\lambda = V_{\rm eff} M_0/\gamma$, where $V_{\rm eff}$ is the effective volume of the magnetic material of the free layer involved in the auto-oscillations, $M_0$ is the saturation magnetization of the free layer, and $\gamma$ is the gyromagnetic ratio. The stationary generation power of the STO is given by $p_0 = (\zeta - 1)/{\zeta + Q})$, where $\zeta = I/I_{\rm th}$ is the supercriticality parameter. The threshold current for auto-oscillations, $I_{\rm th}=\Gamma_0/\sigma$, is determined by the condition by which the linear components of the negative and positive damping terms are equal, $\Gamma_-(0)=\Gamma_+(0)$. In the supercritical region, sufficiently far above the generation threshold, the stationary power $p_0$ of a nonlinear auto-oscillator is found from the condition $\Gamma_+(p_0)=\Gamma_-(p_0)$.

To determine the generation linewidth in the far supercritical regime, it is sufficient to linearize Eq.~(\ref{model}) by examining the time evolution of the power fluctuations $\delta p(t) = p(t) - p_0$ about the steady-state trajectory, as $\delta p$ is much smaller than the stationary value $p_0$. By linearizing Eq.~(\ref{model}) near the point $p=p_0$, we find the following two equations for the power $\delta p$ and phase $\phi$ fluctuations of the auto-oscillations,
\begin{subequations}
\label{eq:linearEOM}
	\begin{eqnarray}
	\frac{d \delta p}{dt} + 2 \Gamma_{\rm p} \delta p &=& 2 \sqrt{p_0} \, {\rm Re}	[\tilde{f}_n(t)], \\
	\frac{d \phi}{d t} + \omega(p_0) &=& \frac{1}{\sqrt{p_0}} {\rm Im}[\tilde{f}_n(t)] - N \,	\delta p,
\end{eqnarray}
\end{subequations}
where $\tilde{f}_n(t) = f_n(t) e^{-i \phi(t)}$ is a stochastic process with the same properties as $f_n(t)$, $N= \partial_p \omega(p)$ is the coefficient of the nonlinear frequency shift,
\begin{equation}
\Gamma_{\rm p} = \partial_p (\Gamma_+ - \Gamma_-)p_0 = G_{\rm eff}p_0,
\label{eq:Gamma}
\end{equation} 
is the damping rate for the power fluctuations, and all the derivatives are calculated at $p=p_0$. It is immediately evident that the nonlinearity $N$ leads to a coupling between the phase and amplitude fluctuations, which generates an additional noise source $-N \delta p$ for the phase variable $\phi$. The physical mechanism behind this additional noise term can be understood in terms of an additional frequency modulation $\omega (p(t)) \approx \omega(p_0) + N \, \delta p$ which arises from the power fluctuations $\delta p (t)$ and the frequency nonlinearity. This additional noise term can be considered as an inhomogeneous broadening of the oscillator linewidth: oscillators with different powers (due to thermal fluctuations) $p$ have different oscillation frequencies $\omega(p)$.

Since the stochastic system (\ref{eq:linearEOM}) is a linear system of equations and the noise $\tilde{f}_n(t)$ is a Gaussian process, both $\delta p (t)$ and $\phi(t)$ are also Gaussian processes for which the complete set of statistical characteristics can be easily obtained. As such, we find a vanishing mean value for the power fluctuations, $ \langle \delta p(t) \rangle = 0$, with the two-time correlation function of the form,
\begin{equation}
\langle \delta p(t) \delta p(t')  \rangle = p_0 \, \eta(p_0) \, \frac{\Gamma_+(p_0)}{\Gamma_{\rm p}} e^{-2 \Gamma_{\rm p}|t-t'|}.
\end{equation}
The mean value of the phase $\phi(t)$ (for a given initial value $\phi_0$ at $t=0$) is given by
%
$\langle \phi(t) \rangle = \phi_0 - \omega(p_0) t$,
%
and the variance of the phase fluctuations is given by
\begin{equation}
\Delta\phi^2(t) = 2\Delta\omega_0 \left[ (1+\nu^2)|t| - \nu^2 \frac{1-e^{-2 \Gamma_{\rm p} |t|}}{2 \Gamma_{\rm p}} \right],
\label{eq:phasevar}
\end{equation}
where
\begin{equation}
\nu = \frac{\partial_p\omega}{\partial_p (\Gamma_+ - \Gamma_-)}= \frac {N}{G_{\rm eff}}
\label{eq:nu}
\end{equation}
is the normalized dimensionless nonlinear frequency shift coefficient which characterizes the relative influence of the frequency nonlinearity $N$ compared to the nonlinearity of total effective dissipation $G_{\rm eff}$ on the auto-oscillator behavior, and $2\Delta \omega_0$ is the generation linewidth of a \emph{linear} ($\nu = 0$) auto-oscillator,
\begin{equation}
2\Delta \omega_0 = \Gamma_+ \frac{k_B T}{\mathcal{E}(p_0)}.
\end{equation}

In stark contrast to the case of a linear oscillator for which $|\nu| \ll 1$, the phase variance $\Delta\phi^2(t)$ for an auto-oscillator with strong frequency nonlinearity $\nu \gg 1$ has a nonlinear dependence on the time interval $|t|$. This nonlinear dependence results from the finite correlation time $\tau = 1/\Gamma_{\rm p}$ of power fluctuations and by the additional nonlinear source of colored noise $-N \delta p(t)$ in the equation of motion for the phase variable (\ref{eq:linearEOM}) as mentioned previously. Only for time intervals $|t|$ that are much larger than the correlation time of the power fluctuations $1/\Gamma_{\rm p}$ does the phase variance vary linearly as a function of time. Thus, the power spectrum of a nonlinear auto-oscillator is non-Lorentzian in general.

The power spectrum $\mathcal S(\Omega)$ of the  nonlinear auto-oscillator (\ref{model}) in the presence of thermal noise,
\begin{equation}
\mathcal{S}(\Omega) = \int d\tau \; \mathcal{K}(\tau) e^{i \Omega \tau},
\end{equation}
is, up to a constant of proportionality, the Fourier transform of the two-time correlation function of the oscillator variable $c(t)$,
\begin{equation}
\mathcal{K}(\tau) = \langle c(t+\tau)c^*(t) \rangle.
\end{equation}
In the supercritical regime in which the oscillator power is stable and phase fluctuations dominate, we can consider simply the phase noise contributions to the power spectrum. In this case, the autocorrelation function corresponding to the random phase can be written as
\begin{equation}
\mathcal{K}(t) = p_0 \, e^{i \langle \phi(t) - \phi(0) \rangle} \exp\left[-\Delta\phi^2(t)/2\right].
\label{eq:autocorr}
\end{equation}
For an STO, the power spectrum measured in a typical experiment corresponds to the Fourier transform of the autocorrelation function of the variable \textit{voltage} that arises from variations in the giant or tunnel magnetoresistance. The relation between the autocorrelation function of the STO voltage and the  autocorrelation function $\mathcal{K}(\tau)$ of the spin-wave auto-oscillation has been derived elsewhere (e.g., see Eq.~(10) in Ref.~\onlinecite{Kim:PRB:2006}).

The form of the phase variance in (\ref{eq:phasevar}) does not permit a simple analytical expression for the Fourier transform of the autocorrelation function (\ref{eq:autocorr}) to be obtained in the general case. However, if the temperature $T$ and, respectively, generation linewidth $\Delta \omega$ are sufficiently small, $\Delta \omega \ll \Gamma_{\rm p}$, one can neglect the exponential factor in (\ref{eq:phasevar}) at the characteristic decoherence time scale $t \sim 1/\Delta \omega$,
\begin{equation}
\Delta\phi^2(t) \approx 2 \Delta\omega_0 (1 + \nu^2)|t| - \frac{\nu^2}{2 \Gamma_{\rm p}}.
\end{equation}
In this low-temperature limit the autocorrelation function $\mathcal{K}(t)$ is a simple exponential function, which leads to a power spectrum $\mathcal{S}(\Omega)$ with a Lorentzian shape and a full width at half maximum of~\cite{Kim:PRL:2008}
\begin{equation}
2\Delta\omega_{\rm LT} = 2\Delta\omega_0 (1+\nu^2) = \Gamma_+(p_0) \frac{k_B T}{\mathcal{E}(p_0)}(1+\nu^2).
\label{eq:linearT}
\end{equation}
In this limit, the frequency nonlinearity leads to an increase in the generation linewidth by a factor of $(1+\nu^2)$ with a linear dependence on the temperature.  One immediately observes that the condition $\Delta \omega \ll \Gamma_{\rm p}$ used to derive Eq.~(\ref{eq:linearT}) can be rewritten as a condition for the temperature
\begin{equation}
k_B T \ll \left( \frac{\Gamma_{\rm p}}{\Gamma_+(p_0)} \right) \frac{\mathcal{E}(p_0)}{1+\nu^2},
\label{eq:linearcond}
\end{equation}
or expressed as a condition for the relative magnitude of the power fluctuations, $\Delta p / p_0 \ll (1+\nu^2)^{-1/2}$. Thus, (\ref{eq:linearT}) represents the low-temperature asymptotics of the generation linewidth for a nonlinear auto-oscillator. Estimates made using Eq.~(\ref{eq:linearT}) for typical parameters of spin-torque oscillators (permalloy circular nanopillar of the radius $R_c = 50$ nm and thickness of the free layer $L=5$ nm, generation frequency $\omega(p_0)/2\pi \simeq 30$ GHz) show that (\ref{eq:linearT}) is quantitatively correct for temperatures in the range $T \leq 10-100$ K, depending on the supercriticality $\zeta$ (see, e.g., Figs. 2a and 2b in Ref.~\onlinecite{Kim:PRL:2008}).

A simple analytical solution for the linewidth can also be obtained for the opposite limit of high temperatures in which the correlation time of the power fluctuations ($1/\Gamma_{\rm p}$) is much longer than the coherence time of the oscillator ($\sim 1/\Delta \omega$). In this limit, the exponential function in (\ref{eq:phasevar}) can be expanded in a Taylor series to give
\begin{equation}
\Delta\phi^2(t) \approx 2\Delta\omega_0 (|t| + \nu^2 \Gamma_{\rm p} t^2).
\end{equation}
For large frequency nonlinearities, $\nu \gg 1$, one can neglect the linear term in $|t|$ and keep only the quadratic term in the Taylor series expansion, which leads to a \emph{Gaussian} power spectrum,
\begin{equation}
\mathcal{S}(\Omega) \sim \exp\left[ \frac{(\omega(p_0) - \Omega)^2}{2 \Delta\omega_{\rm HT}^2}   \right],
\end{equation}
where the characteristic linewidth is defined by
\begin{equation}
\Delta\omega_{\rm HT} = |\nu| \sqrt{\Gamma_+(p_0) \Gamma_{\rm p}} \sqrt{\frac{k_B T}{\mathcal{E}(p_0)}}.
\label{eq:sqrtT}
\end{equation}
It is interesting to note that the generation linewidth in this regime is proportional to $\sqrt{T}$, in contrast to the linear temperature dependence (\ref{eq:linearT}) of the linewidth in the low-temperature regime. Such a temperature dependence has been originally proposed in Ref.~\onlinecite{Sankey:PRB:2005} based on macrospin simulations of an STO. In the context of the present work, this linewidth behavior can be understood as an apparent \emph{inhomogeneous broadening} by power fluctuations which are translated into a modulation of the oscillation frequency through the frequency nonlinearity. The expression (\ref{eq:sqrtT}) for the generation linewidth is valid in the temperature range
\begin{equation}
\left( \frac{\Gamma_{\rm p}}{\Gamma_+(p_0)}\right)\frac{\mathcal{E}(p_0)}{\nu^2} \ll k_B T \ll \left( \frac{\Gamma_{\rm p}}{\Gamma_+(p_0)}\right) \mathcal{E}(p_0),
\label{eq:sqrtcond}
\end{equation}
which can also be rewritten in terms of the range for the power fluctuations as  $1/|\nu| \ll \Delta p / p_0 \ll 1$.

To study the transition between the low and high temperature asymptotic behavior, we performed numerical calculations of the Fourier transform of autocorrelation function (\ref{eq:autocorr}) using the phase variance given by the general expression (\ref{eq:phasevar}).
\begin{figure}
\includegraphics[width=7.5cm]{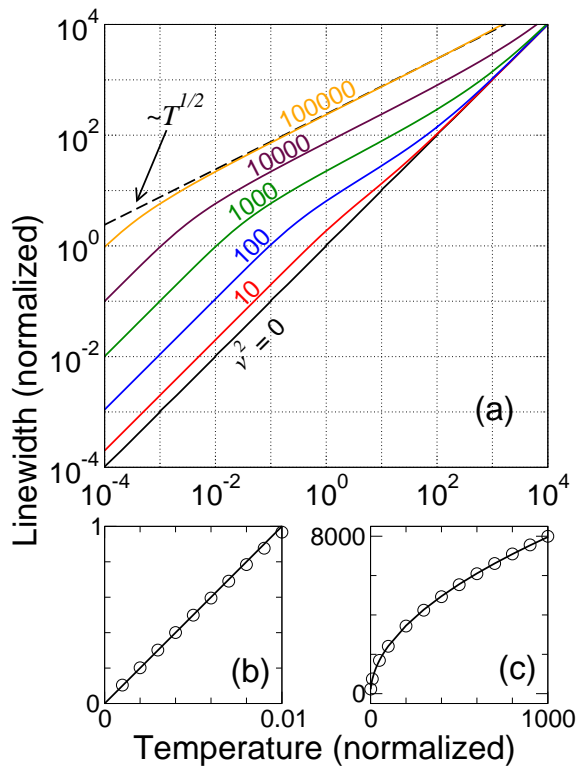}
\caption{\label{fig:linewidth_numerical} (Color online) Normalized linewidth, $\Delta \omega' = \Delta \omega/\Gamma_{\rm p}$, as a function of the dimensionless temperature $T'$. (a) Linewidth variations for different values of the frequency nonlinearity $\nu$.  (b) Low temperature asymptotics for $\nu^2=100$ (circles are numerical data, solid line is fit using the linear function $\Delta f' = 100 T'$). (c) High temperature asymptotics for $\nu = 10^4$
[circles are numerical data, solid line is a fit using the function $\Delta f' = 222 \, (T')^{0.518}$].}
\end{figure}
For the numerical work it is convenient to use the dimensionless definitions of time, $t' = \Gamma_{\rm p} t$, and temperature, $T' = \left[ (\Gamma_+(p_0) k_B)/(\Gamma_{\rm p} \mathcal{E}(p_0))  \right] T$.
%
%
With these scaled variables, the phase variance in (\ref{eq:phasevar}) can be simplified to 
\begin{equation}
\Delta\phi^2(t') = T' \left[ (1+\nu^2)|t'| - (\nu^2/2) (1-e^{2|t'|})  \right].
\end{equation}
The results of the numerical calculations of the generation linewidth (normalized by $\Gamma_{\rm p}$), for several values of the normalized nonlinearity coefficient $\nu$, as a function of the dimensionless temperature $T'$ over 8 orders of magnitude are shown in Fig.~\ref{fig:linewidth_numerical}. The linewidth is determined from the full width at half maximum of the power spectrum, without fitting to any particular line shape. 

In the absence of nonlinearity the linewidth is simply proportional to the temperature, and in our reduced units, there is a one-to-one correspondence between the two quantities. As the nonlinearity $\nu$ is progressively increased, the temperature range over which the linear behavior holds decreases over the temperature range studied, or, in other words, the linear regime is pushed toward lower temperatures, as indicated by (\ref{eq:linearcond}). In the low temperature regime the linewidth dependence on $\nu$ is consistent with Eq.~(\ref{eq:linearT}), i.e. the linewidth increases as $(1+\nu^2)$, as expected. As the temperature is increased, a gradual transition toward the square-root temperature dependence is observed. Over the temperature range studied, this transition occurs more rapidly for the larger values of $\nu$, as expected from (\ref{eq:sqrtcond}). The low and high temperature limits are shown explicitly in Figs.~\ref{fig:linewidth_numerical}b and \ref{fig:linewidth_numerical}c, respectively.

It is possible to test our theory directly in experiment by tuning the nonlinearity coefficient $\nu$ over a reasonably wide interval. For STOs, this can be achieved by varying the orientation of the out-of-plane external bias magnetic field relative to the plane of the ``free" layer for magnetic nano-contact system,~\cite{Rippard:PRB:2004} or by varying the orientation of the in-plane bias magnetic field relative to the ``easy" anisotropy axis in an anisotropic in-plane magnetized magnetic nanopillar.~\cite{Thadani:2008} In both cases, the variation in $\nu$ is due to different dynamic dipolar fields generated as a result of changes in the magnetization precession axis.~\cite{Slavin:ITM:2005} We would like to stress that any quantitative comparison between our theory and experiment also requires detailed knowledge of the temperature dependence of all the key magnetic parameters of the STO, such as oscillation frequency, shape anisotropy, and relaxation rate, so that the supercriticality can be determined accurately for a given value of the bias current. One way to overcome such difficulties is to drive the STO with an external noisy microwave magnetic field at {\em fixed} temperature. In this case the effective noise temperature is proportional to the power $P_n$ of microwave noise source and the dependence $\Delta\omega(P_n)$ should be qualitatively the same as the temperature dependence $\Delta\omega(T)$, in particular, $\Delta\omega \sim P_n$ for small $P_n$ and $\Delta\omega \sim \sqrt{P_n}$ for relatively large $P_n$.

In summary, we have developed a theory for the temperature dependence of a generation linewidth of an auto-oscillator with a large nonlinear frequency shift. At low temperatures, the nonlinearity leads to the renormalization of the phase noise, which results in a Lorentzian power spectrum and a linear temperature dependence. At high temperatures, the nonlinearity leads to an apparent inhomogeneous broadening of the spectral line, giving rise to a Gaussian line shape and a square-root dependence on the temperature. 

\begin{acknowledgments}
This work was in part supported by the MURI grant W911NF-04-1-0247 from the US Army Research Office, the contract W56HZV-07-P-L612 from the U.S. Army TARDEC, RDECOM, the grant ECCS-0653901 from the National Science Foundation of the USA, the Oakland University Foundation, and the European Communities program IST under Contract No. IST-016939 TUNAMOS.
\end{acknowledgments}

\bibliography{articles}
\end{document}